\RequirePackage[2020-02-02]{latexrelease}
\documentclass[twocolumn,amsmath,amssymb,fleqn,prb]{revtex4}

\usepackage{cancel}
\usepackage{graphicx}%
\usepackage{bm}
\usepackage{subfigure}

\begin{document}

\title{Can 1957 cold-fusion explain 1989 cold-fusion?}
\author{S. Di Matteo$^1$}
\affiliation{$^1$ Univ Rennes, CNRS, IPR (Institut de Physique de Rennes) - UMR 6251, F-35708 Rennes, France}

\date{\today}

\begin{abstract} 
The possibility that muon-catalyzed nuclear fusion at ambient temperature takes place in deuterated metals is analyzed theoretically. It is suggested that the muon-catalyzed deuterium-deuterium ($dd$) or deuterium-tritium ($dt$) fusion rate, experimentally observed in liquid deuterium, increases in a PdD crystal. The main reason is that, in the palladium crystal matrix, deuterium diffuses in ionic ($d^+$), rather than molecular ($D_2$), form, thereby favoring the rate of (d$\mu$d)$^+$ formation. The slightly enhanced deuterium density and the possibly reduced muon sticking probability (compared to liquid hydrogen) point to an increase of the fusion-rate, unfortunately counterbalanced by the probability of $\mu^-$ capture by the Pd nucleus, whose value should be evaluated by a dedicated experiment. 

If this mechanism were possible, we advance the hypothesis that 1989 {\it cold fusion} might have been randomly triggered by cosmic muons: an order-of-magnitude analysis shows that, under special conditions, a mW power per $\mu^-$ might be produced. Moreover, recent experimental results showing that neutrons can sustain deuterium fusion in out-of-equilibrium conditions might explain the remaining power generation. However, reliable quantitative figures are difficult to obtain theoretically, and an experiment is suggested, to be conducted at muon-beam sources, to verify or invalidate this idea.

\end{abstract}

\maketitle

\section{Introduction}

Two different phenomena were named {\it cold fusion} in the scientific literature at different epochs. The first, in 1957, concerned the muon-driven deuterium fusion at ambient temperature\cite{alvarez,jackson1,jackson2,prlexp1987,aipconfproc,petitjean} and the second, in 1989, was associated to an eventual excess-power production in  electrolithic cells when a palladium electrode was charged with deuterium\cite{pons1,pons2,petrasso,salamon}. Though neither phenomenon could lead to the energy production hoped for, their history is profoundly different. 

The 1957 muon-driven deuterium fusion was correctly interpreted by J.D. Jackson\cite{jackson1,jackson2} in terms of the available theory of nuclear reactions and led to a deeper understanding of muon interactions with hydrogen isotopes\cite{vesman,ponomarev1,ponomarev2}, all confirmed experimentally\cite{petitjean}.
In particular, it became clear that the limitations of the technique to reach the break-even point for technological applications were intrinsic to the process (see Section II) and thereby (probably) insurmountable. 

The 1989 cold fusion, instead, was plagued since its beginning by unusual and unconvincing procedures of scientific communications\cite{petrasso,salamon} and the authors were unable to provide at least one of the two {\it tenet} that are required for the general acceptance of a new - and potentially important - phenomenon: {\it a)} reproductibility of experiments; {\it b)} an acceptable theoretical explanation\cite{note1}. Indeed, on one side, in spite of some experiments confirming the excess power production, many others did not. On the other side, no acceptable theoretical phenomenology was found, based on the available nuclear theory, that did not appeal to some {\it miracles}, like\cite{scara} the raise of the branching ratio of $dd$ fusion to $^4$He+$\gamma$+24 MeV, from the accepted value $10^{-6}$ to almost 1. To fix the ideas, state-of-the-art calculations of $dd$ fusion in Pd crystal matrix, to any of the possible reaction products (e.g., Ref. [\onlinecite{rmp-jap}], p. 284-285) at ambient temperature, lead to less than two nuclear reactions per cm$^3$ {\it per year}. 

In spite of the absence of any sound theoretical explanation, it might be fair to underline that several independent experimental verifications have been conducted around the world and that, given the random successful results\cite{scara,cfyes1,cfyes2,cfno1,cfno2,cfno3}, if we do not believe the idea of a large-scale fraude, we should rather point to an objective phenomenon characterized by uncontrolled, casual, positive outcomes.
In the present paper we adopt such an hypothesis and look for a known theoretical framework that explains the phenomenology of 1989 cold fusion. As, today, the only known way to let two deuterons approach up to the fusion-reaction point at ambient temperature is provided by muons (electron screening alone is unefficient\cite{rmp-jap}, though for Pd it is as high\cite{terrasi} as 800 eV), the most natural answer to this question leads to the suggestive hypothesis that the two phenomena, the 1989 and the 1957 {\it cold fusions}, could be related. Though apparently no muons were present in 1989 and related experiments, we analyze the possibility that cosmic muons could have triggered deuterium fusion: under special circumstances, one muon can trigger more than a hundred nuclear-fusion $dt$ reactions and more than 10 $dd$ reactions (see next sections). This might in part explain the randomness associated to 1989 {\it cold fusion}, as well as the fact that the same experimental apparatus could lead to power production at the INFN laboratory in Frascati and not at the INFN laboratory in Gran Sasso\cite{scara} (where the muon flux lowers by about a factor $10^{-4}$). This also implies that 1989 cold fusion is a non-equilibrium phenomenon, triggered by an external energy source - in our hypothesis the muonic part of cosmic rays.

In short, the suggested mechanism to explain 1989 {\it cold fusion} goes as follows (the origin of the numerical values cited in this section is explained in the sections II, III and IV): a low-energy (momentum less than few MeV/c) cosmic negative muon ($\mu^-$) stops at a deuterium\cite{diffdD} (d$^+$) site in PdD, forming a muonic hydrogen atom. The muonic heavy-hydrogen atom diffuses (in a typical time $t_d\simeq 10^{-10}$ s, see Section III.A) from one octahedral interstitial site to the next, where it forms a muonic $^{\mu}\!D_2^+$ molecule in a time less than $t_d$. The two D nuclei in the muonic $^{\mu}\!D_2^+$ molecule have an average distance of $\simeq 270$ fm and the vibrational motion of the molecule at ambient temperature can bring the two nuclei at a relative distance of less than 5 fm, where strong interactions can lead to deuterium fusion through the two channels: 

\vspace{-0.5cm}
\begin{subequations} 
\begin{align}
&  d^+ + d^+ + \mu^- & \hspace{-0.5cm} &  \longrightarrow  \hspace{0.3cm} p^+ + t^+ + \mu^- (+ Q_1) \label{ddfus1} \\
& d^+ + d^+ + \mu^- & \hspace{-0.5cm}  & \longrightarrow \hspace{0.2cm} ^3\hspace{-0.02cm}{\rm He}^{++} + n^0 + \mu^- (+ Q_2) \label{ddfus2}
\end{align}
\label{ddfus}
\end{subequations}
\vspace{-0.5cm}

\noindent with $Q_1\simeq 4.0$ MeV and $Q_2\simeq 3.3$ MeV. The two channels are roughly equiprobable, but some temperature dependence of their relative weight was experimentally found\cite{petitjean}.
We remark that the presence of the muon may lead to a different energy distribution of the nuclear-reaction outcomes, that are no more constrained by the two-particle fixed-energy distribution (i.e., the neutron of reaction (\ref{ddfus2}) does not need to have the fixed energy of 2.475 MeV, as some energy and impulse is taken by the outgoing muon). We also remark that reducing temperature suppresses the vibrational amplitude and therefore fusion probability.

However, a single muon from cosmic rays, even if triggering about 10 or 100 fusion reactions (for $dd$ and $dt$ cases of equations (\ref{ddfus}) and (\ref{dtfus}), respectively\cite{aipconfproc,petitjean}), cannot explain the magnitude of the excess power claimed in some experiments (a few watts during several minutes): with the present knowledge of 1957 cold fusion applied to a PdD crystal, one muon catalyzing $dd$ fusion might only trigger a milliwatt power, for fractions of $\mu$s (see sections II and III).
As the cosmic muon flux at the typical size of a PdD electrode is around one $\mu^-$ per minute\cite{notamu}, cosmic muons cannot explain, alone, the claimed results. In a sense, this is conceivable - otherwise we might wonder why cosmic muons did not induce the same power production in deuterated plasmas. In the picture that we propose, cosmic muons only act as triggers - they locally heat the deuterium fuel - and then a new phenomenon, typical of condensed matter, gains the upper hand.
This phenomenon has been demonstrated only very recently\cite{prc1,prc2}: when some energy is provided to deuterated metallic crystals TiD$_2$ and ErD, in the form of 2 MeV photons, then deuterium fusion reactions can take place. In fact, deuterium nuclei acquire sufficient kinetic energy to overcome the Coulomb barrier (slightly reduced by the electron screening typical of condensed-matter environments) through the collisions with the 'hot' neutrons (between 300 KeV and 2 MeV) produced by the breaking of the deuterium isotope by the 2 MeV photons. The key point here is that any mechanism allowing to produce these 'hot' neutrons would work as well to trigger further fusion reactions (i.e., not only the two-MeV photons, but also, e.g., the muon-catalyzed fusion reaction \ref{ddfus2}). However, any theoretical calculation of the efficiency of the 'hot' neutron mechanism to continue the fusion process would be biased by too many unknowns, notably about the creation of defects and deformations of the crystal lattice. For this reason, though in the next sections, on the basis of the known theory, we propose an educated guess about what could have happened, we believe that only a dedicated experiment, to be conducted at muons facilities, can definitely settle the question. 

The present paper is organised as follows : Section II is devoted to the description of the main aspects of 1957 {\it cold fusion} up to the most recent findings\cite{petitjean}. We limit our main analysis to $dd$ fusions (\ref{ddfus}) because our aim is to explain 1989 cold fusion, where mainly $dd$ fuel was used. Yet, $dt$ fusion:

\begin{align}
d^+ + t^+ + \mu^- \longrightarrow  ^4\hspace{-0.02cm}{\rm He}^{++} + n^0 + \mu^- (+ Q_3)
\label{dtfus}
\end{align}

\noindent with $Q_3\simeq 17.6$ MeV, can in principle provide a better efficiency\cite{aipconfproc} and we shall also briefly linger on it in.

In section III we focus on the peculiar features of Pd crystals and their capabilities of absorbing H-atoms (any isotope). The condensed-matter properties of PdD crystals (1:1 ratio of palladium and deuterium) are investigated, in particular those of interest for the 1989 {\it cold fusion}. We shall analyze how the rate of formation of muonic $^{\mu}\!D_2^+$ molecules, how the sticking probability can be changed by the crystal Pd matrix (compared to liquid deuterium) and how cosmic muons might stop in PdD.

Section IV reviews some very recent findings\cite{prc1,prc2} pointing to the role of 'hot' neutrons (hundreths of keV, up to few MeV) to trigger nuclear fusion reactions of the kind (\ref{ddfus}) within a crystal matrix. This section also highlights why the proposed $\mu^-$-triggered nucelar fusion would not work in, e.g., deuterated plasmas. 

In Section V we draw our conclusions, with a focus on the three critical parameters of this work, that, being mainly unknown in the literature and for which no reliable calculations can be performed to the best of my knowledge, have been subjected for the moment to an educated guess. They are: 1) the probability of $\mu^-$ implantation at deuterium octahedral sites (called $1-p_X$ below); 2) the probability of muon reactivation (or $^3\hspace{-0.02cm}{\rm He}^{+}$ muon stripping) in the Pd crystal (called $p_{st}$ in Sections II and III.A); 3) The efficiency of the relatively hot ejected neutrons to maintain the reaction triggered by the muon.

\section{Muon-catalyzed (1957) cold fusion in fluid deuterium: a review}

It is common knowledge that a gas of ionic/atomic/molecular hydrogen isotopes (protium, $p$, deuterium, $d$, tritium, $t$) cannot undergo nuclear fusion at ambient temperature because the low, thermal, kinetic energy does not allow to overcome the repulsive Coulomb barrier of the two $H$ (any of $p$, $d$, $t$) nuclei, as $\simeq 300$ keV of kinetic energy are needed to reach a relative distance of few fm, where strong interactions can lead to nuclei fusion with a non-negligible probability.

Yet, in 1957, L. Alvarez and coworkers\cite{alvarez} found nuclear fusion in a liquid-hydrogen bubble chamber at ambient temperature under $\mu^-$ irradiation, a process that was theoretically quantified by J.D. Jackson a few months later\cite{jackson1} as driven by the catalytic role of negative muons, $\mu^-$. Indeed, negative muons are extremely efficient in screening the repulsive Coulomb barrier of the two $H$ nuclei,\cite{jackson1b} either through the formation of an intermediate muonic $^{\mu}\!H_2^+$ molecule or through direct barrier reduction in the scattering process. In the former case, $H$ nuclei in the muonic $^{\mu}\!H_2^+$ molecule have an average distance of $a_0\frac{m_e^*}{m_{\mu}^*}\simeq 270$ fm, compared to the typical Bohr's radius $a_0\simeq 53$ pm for electronic H$_2$ molecules (here $m_e^*\simeq m_e$ and $m_{\mu}^*\simeq 0.95 m_{\mu}\simeq 195 m_e$ are the effective masses of electronic and muonic deuterium atoms, respectively).  
At ambient temperature, the vibrational motion in the muonic $^{\mu}\!H_2^+$ molecule allows the two nuclei to come sufficiently close (up to few fm), so as to have a non-negligible fusion probability per unit time. After the fusion process takes place, the muon might still be free to trigger further reactions, within its lifetime $\tau_{\mu}\simeq 2.2\cdot 10^{-6}$ s.
To fix the orders of magnitude, we remind that the rate for the reactions (\ref{ddfus1}) and (\ref{ddfus2}) was estimated in Ref. [\onlinecite{jackson1}] to be $\simeq 10^{10}$ s$^{-1}$ (their branching ratio is 50\%) and up to $\simeq 10^{12}$ s$^{-1}$ for the $dt$ fusion reaction of Eq. (\ref{dtfus}). 
Given these values, if we consider for example $dt$ fusion reaction (leading to the maximum power release), at a rate of $10^{12}$ Hz for a time $\tau_{\mu}$, with an energy release of 17.6 MeV per reaction, we end up (ideally) with a maximum power release of $17.6 \cdot 1.6\cdot 10^{-13}\cdot 10^{12}\simeq 3$ W per muon inserted in the deuterium-tritium gas (corresponding to an energy of $3\cdot 2.2\cdot10^{-6}\simeq 6$ $\mu$J), if all reactions were of the $dt$ kind. With the same calculation, the average of reactions (\ref{ddfus1}) and (\ref{ddfus2}) for a pure deuterium gas gives 5 mW (corresponding to an energy of 11 nJ). However, as Jackson found in 1957, we cannot get these values experimentally for two reasons: a) it takes some time (estimated by Jackson $\sim 2\cdot 10^{-8}$ s) to form a muonic $H_2^+$ molecule and b) there is a non-negligible probability (slightly less than $1\%$ for $dt$ reaction, and about $12\%$ for reaction (\ref{ddfus2})) that the $\mu^-$ sticks to the decay ${\rm He}^{++}$ particle and is therefore lost in its role as a catalyst. Jackson's calculations showed that the formation rate of a muonic molecule in liquid hydrogen implied no more than 100 fusion catalyst processes per average muon. Moreover, the sticking probability led the average muon not to trigger more than 100 reactions for $dt$ fusion (instead of $2\cdot 10^6$ of the previous estimate), or no more than 10 for (\ref{ddfus2}) (instead of $2\cdot 10^4$). For these reasons, the power of the previous (ideal) estimates should be multiplied by $\frac{0.5\cdot 10^8 {\rm s}^{-1}}{10^{12} {\rm s}^{-1}}\simeq 5\cdot 10^{-5}$ (for $d-t$ reactions, leading to 150 $\mu$W instead of 3 W) or by $\frac{0.5\cdot 10^8 {\rm s}^{-1}}{10^{10} {\rm s}^{-1}}\simeq 5\cdot 10^{-3}$ (for $d-d$ reactions, leading to $\simeq 25$ $\mu$W instead of 5 mW). In both cases, the reactions would switch off in a time determined by the sticking probability, smaller than the muon lifetime. The corresponding energy production per $\mu^-$ would be $150$ $\mu$W$\cdot 100$ reactions $\cdot 2\cdot10^8$ s$\simeq 300 pJ$ for the $dt$ case and $25$ $\mu$W$\cdot 10$ reactions $\cdot 2\cdot10^8$ s$\simeq 5 pJ$ for the $dd$ case.     
 The energy gain for either of the previous reactions would be less than the energy needed to produce a muon, about 5 GeV$\simeq 800$ pJ. For this reason, $\mu^-$ catalyzed nuclear fusion was not considered worthwhile for energy production at the beginning of 1960s. However, researches continued on the other side of the iron curtain, and soviet scientists discovered a resonant formation channel, strongly increasing\cite{vesman,ponomarev1} the formation rate of muonic $^{\mu}\!H_2^+$ molecules estimated by Jackson. Therefore in the 1980s there was some renewed interest on muon-catalyzed fusion also in the United States, and new experiments\cite{prlexp1987} showed that up to 150 reactions (for $dt$ fusion reaction) could be driven by one $\mu^-$ particle. Theoretical estimates were also readjusted by better estimating shake-off reactivation processes of $\mu^-$ particles\cite{jackson2,aipconfproc}. But, the break-even point (where $\mu^-$-catalyzed reactions becomes useful for energy production) could not be reached and the research in this field was practically abandonned, in spite of more recent hints\cite{japreactivation} to overcome muon sticking. The most detailed, recent, experimental analysis about the $dd$-processes of Eqs. (\ref{ddfus1}) and  (\ref{ddfus2}) is reported in Ref. [\onlinecite{petitjean}], together with the state-of-the-art theoretical explanations.

We close this short review on 1957 cold fusion by remarking that, as the muon lifetime is fixed by nature, as well as the rate of reactions (\ref{ddfus}), the only critical parameters that might have allowed a higher reaction rate are {\it a)} the formation rate of the muon molecule, i.e., the probability to get two deuterium ions within few fm (it should increase) and {\it b)} the sticking probability of $\mu^-$ on ${\rm He}^{++}$ (it should decrease). For what the sticking probability, $p_{st}$, is concerned, we remind that it can be conceptually described as the product $p_{st}=p_{is}(1-p_{rs})$ of an initial sticking probability, $p_{is}$, corresponding to the probability of formation of a $\mu{\rm He}^+$ molecule after the fusion reactions (\ref{ddfus2}) or (\ref{dtfus}) and a reactivation factor, $p_{rs}$, measuring the strenght of the shake-off processes that may strip the muon off the He ion, and therefore reduce the value of $p_{st}$. Whereas $p_{is}$ depends on a nuclear process that we cannot change, $p_{rs}$ depends on the ionization rate of the muonic ${\rm He}_{\mu}^{+}$ molecule, that might be increased, for example by increasing the sample density. Indeed, in all experiments performed up to now, the H$_2$ fuel was fluid and the improvement of both {\it a)} and {\it b)} critical parameters was mainly looked for by increasing the density of the H$_2$ gas. 
We can express the above considerations as follows: the number of fusion reactions $N_R$ is given by the ratio of two rates, the rate of the muon-fusion cycle, $f_c$, and the rate of muon losses, $f_l$ (forcing the muon to exit the cycle). The rate of the muon-fusion cycle, mainly determined by the rate of $^{\mu}\!H_2^+$ molecule formation, was estimated by Jackson as $f_c\simeq 10^8$ s$^{-1}$. Among the losses, the finite lifetime of the $\mu^-$, $f_{\mu}\simeq 4.55\cdot 10^5$ s$^{-1}$ is not the critical parameter, rather the losses determined by the sticking probability, $f_{st}\simeq p_{st}f_c$, play a major role (see Eq. (\ref{mucycle}), with $p_{st}\simeq 0.12$ for $dd$ fusion\cite{jackson1}. If we include in the expression for $N_R$ any other losses, $f_X=p_Xf_c$ (for the moment $p_X=0$), we get the final formula for the number of fusion reactions $N_R$ obtained by one muon:
\vspace{-0.1cm}
\begin{align}
N_R = \frac{f_c}{f_l}=\frac{f_c}{f_{\mu}+f_{st}+f_X}=\frac{1}{\frac{f_{\mu}}{f_c}+p_{st}}\simeq\frac{1}{p_{st}}\simeq 8
\label{mucycle}
\end{align}
\vspace{-0.3cm}

\noindent where it is clear that, given the above numbers, $p_{st}$ plays a major role compared to $f_{\mu}$ (we remind that for $dt$ fusion, $p_{st}\simeq 0.007$, so that\cite{petitjean} $N_R\simeq 150$).
As we shall see in section III, we argue that the enhanced density determined by Pd ions in PdD increases the number of shake-off processes thereby reducing the effective sticking probability $p_{st}$. Moreover, we show that the molecule formation rate $f_c$ is also increased by the fact that deuterium enters Pd in the ionic $d^+$ form and not as $D_2$. However, this is couterbalanced by a new form of energy loss, as the muon can be captured also by the Pd nucleus ($p_X\neq 0$) and therefore be definitively lost. A detailed Table is presented in Section V.

\section{Condensed-matter properties of deuterated palladium crystals}

\begin{figure}[ht]
\centering
\hspace{-0.2cm}\includegraphics[width=0.5\textwidth]{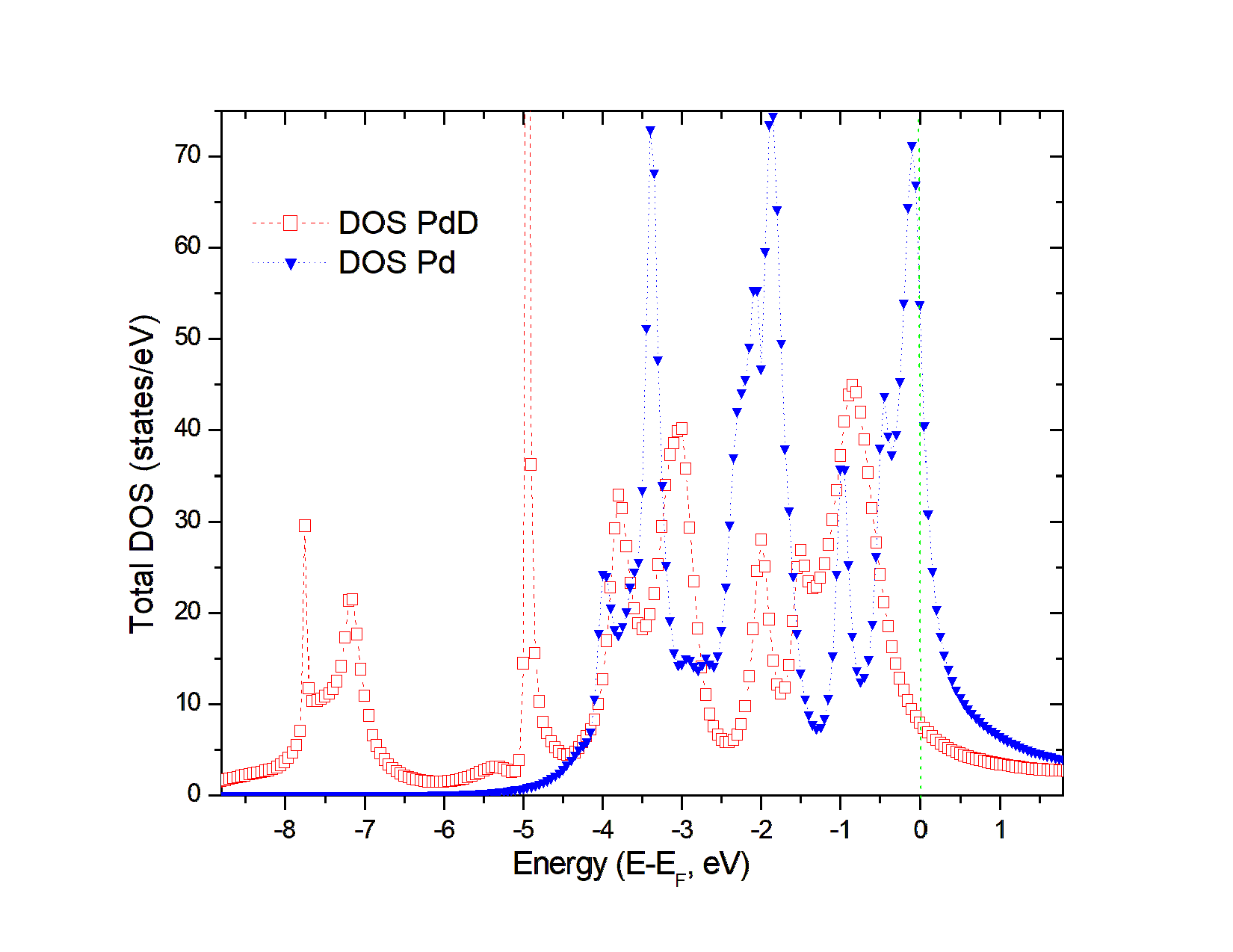}
\caption{Total electronic density of states for Pd and PdH. Energy is measured from the Fermi level. See text.}
\label{fig1} 
\end{figure}

It is textbook\cite{kittel} knowledge in condensed matter that palladium cubic crystal reduces its electronic energy by absorbing hydrogen,\cite{hydrogen} $H$ up to $\simeq 67\%$ of the number\cite{nmr} of palladium atoms. By entering the Pd crystal, the $H_2$ molecules split in two $H$ atoms that, in a first approximation, ionize as $H^+$ by releasing electrons that fill the $4d$ partially empty Pd band, thereby reducing the total energy of the system: the energy with ionized hydrogen inside the palladium crystal is less than the energy of the palladium crystal surrounded by the hydrogen gas by about 0.6 eV per unit cell.\cite{totalenergy} The detailed electronic density of states has been studied more than 40 years ago.\cite{papacon} We performed new calculations through self-consistent finite-difference calculations (cluster radius 5.4 \AA) with the fdmnes program\cite{fdmnes} and summarized the main findings, basically corresponding to those of Ref. [\onlinecite{papacon}], in Fig. \ref{fig1}: electron density originating from 1s H orbitals (the low-energy states lying between -8 and -7 eV in Fig. \ref{fig1}, hybridized with $e_g$ Pd states) leads to a bigger delocalization of the total electron density (more than $\simeq 8$ eV bandwidth for PdH vs. $\simeq 6$ eV bandwidth for Pd), it reduces the total energy per unit cell by about 0.6 eV (numerical integration of the electron DOS) and also strongly suppresses the electron density of states at the Fermi level, as confirmed by magnetic\cite{nmr} and resistivity\cite{resistivity} measurements.

\begin{figure}[ht]
\centering
\includegraphics[width=0.4\textwidth]{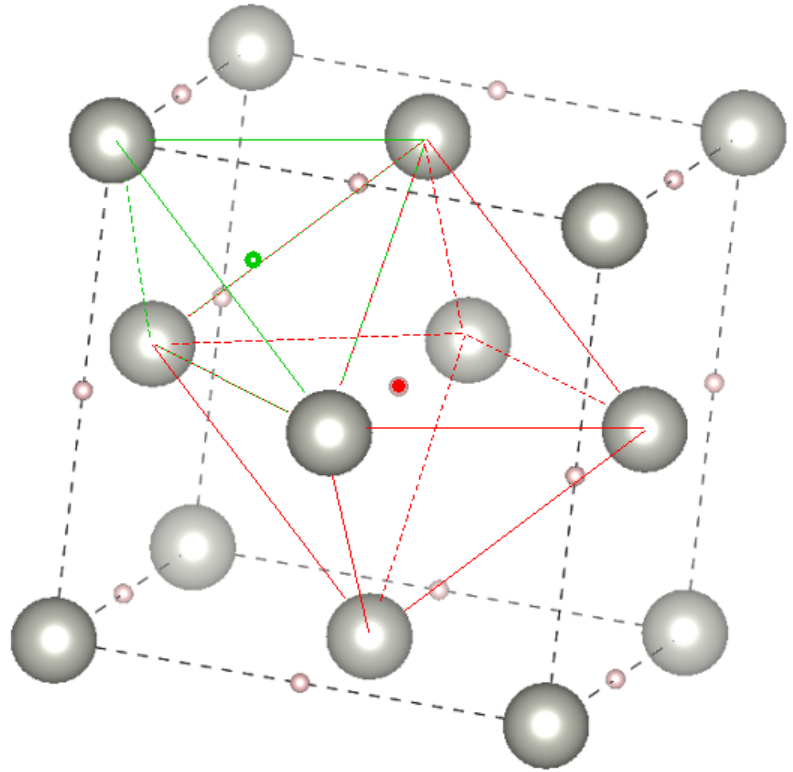}
\caption{Crystal structure of PdH (Pd at ($0$,$0$,$0$) and equivalent positions, H at ($\frac{1}{2}$,$\frac{1}{2}$,$\frac{1}{2}$) and equivalent positions. One of the octahedral interstitial sites of the Pd metal, occupied by H-atoms, is highlighted in red and one of the tetrahedral sites in green.}
\label{fig2} 
\end{figure}

Figure \ref{fig2} shows the cubic PdH crystal structure\cite{neutrondiff} (N. 225 International tables, $m\overline{3} m$ symmetry), with lattice parameter $a\simeq 4.08$ \AA, highlighting the octahedral sites (Wyckoff number 8a) where Pd and H$^+$ ions are located\cite{neutrondiff,svare}, as well as one tetrahedral site (Wyckoff number 16c). As stated above, Pd crystal in an hydrogen atmosphere naturally absorbs $H$ ions in octahedral sites up to $\simeq 60\%$ of the number of Pd atoms.\cite{totalenergy} Yet, this is not the highest absorption limit, that can be pushed up to 100$\%$ and more by, e.g., electrochemical methods. When $100 \%$ filling is obtained (the so called $\beta$-phase), all octahedral sites are filled by hydrogen ions,\cite{neutrondiff} whereas some tetrahedral sites are necessarily filled when\cite{russians} $133\%$ or more\cite{cfno3} is reached. We remark that experimental values of the heat of formation\cite{kuji} for PdH$_x$ from Pd(s) and $\frac{x}{2}$H$_2$(g) indicate a minimum of about $-0.8$ eV for $x\simeq 0.6$ and still a negative value for $x=1$ (though fractions of eV, in competition with temperature in free energy). Self-consistent augmented plane-wave calculations with the Hedin-Lunqvist exchange in the local density approximation,\cite{totalenergy} in comparison, lead to a minimum value of $-0.8$ eV for $x\simeq 0.8$ and a value of $-0.6$ eV for $x=1$. In spite of the quantitative disagreement, both results point to a lower energy for PdH with respect to the separate Pd crystal and H$_2$ gas.

We finally remind the reason why electrochemical loading is more efficient than gas loading at high pressure: the authors of Ref. [\onlinecite{nasa}] could only reach $x=0.9$ through $H_2$ gas loading at 1000 atmospheres, whereas up to $x=0.97$ could be reached through electrochemical methods. As the pressure in the electrochemical cell does not reach 1000 atmospheres, it is obvious that another parameter plays a relevant role to favour hydrogen absorption in the case of electrochemical loading vs. gas loading. This parameter is related to the characteristics of the adsorbing surface that is such as to allow more easily hydrogen atoms to enter than molecules. Indeed, it was convincingly shown in Ref. [\onlinecite{prl3vuoti}] that H$_2$ molecules can be adsorbed by surface Pd atoms only when neighboring octahedral sites of these surface Pd atoms are not filled, so that the adsorbed H$_2$ molecule can split off in two H atoms and both can enter through two (empty) octahedral sites about the Pd ion, in this way decreasing the total energy. For this reason, loading from the gas above a given concentration is statistically more difficult, because it requires (e.g., for the Pd(111) surface studied in Ref. [\onlinecite{prl3vuoti}]) the availability of surface Pd atoms with two empty nearest-neighbours octahedral sites of the three available, vs. only one empty nearest neighbour for electrochemical loading, where hydrogen reaches the Pd surface already in ionic form.

\subsection{Formation rate of muonic $^{\mu}\!D_2^+$ molecules: evaluation of $f_c$ in Equation(\ref{mucycle}).}

$H^+$ ions can diffuse (or hop coherently in some cases\cite{cfno3}) from one octahedral site to another, either directly\cite{svare} or through the tetrahedral sites\cite{cfno3}, depending on the filling\cite{note1b} and on temperature. Experimentally, the typical diffusion rate at low filling and ambient temperature\cite{svare} is around $10^{10}$ s$^{-1}$. This value is obtained by fitting quasielastic neutron scattering, nuclear magnetic resonance and diffusion experiments to a simple tunneling model within the octahedral sites. In the limit $T\rightarrow \infty$, all curves converge to $\tau_{\infty} \simeq 2\cdot 10^{-14}$ s.
Theoretical ab-initio calculations\cite{zunger} show that the energy barrier from octahedral to octahedral sites is less than 0.4 eV (or even 0.23 eV, according to Ref. [\onlinecite{jpsj}], as inferred from diffusivity data) and a direct application of the Heisenberg uncertainty relations for this value gives a transition time $\Delta t \simeq \frac{\hbar}{\Delta E} \simeq 10^{-14} {\rm s} \sim \tau_{\infty}$. For comparison, the energy barrier from the octahedral to the tetrahedral site is about\cite{zunger} 0.3 eV, and yet the double tunnelling (octahedral-tetrahedral-octahedral) through the lower barrier is not advantaged compared to the direct octahedral-octahedral tunnelling\cite{svare}. We remind that these values of the diffusion time between nearest-neighbor octahedral sites correspond to experiments performed in PdH$_x$ for $x \leq 0.7$.  
The only value found in the literature concerning a higher value of $x$ (namely, $x=1.1$) is a molecular-dynamics simulation\cite{cfno3} pointing to a diffusion constant $D_c\simeq 10^{-10}$ m$^2$/s at the temperature of 500 K, leading to an extrapolation at ambient temperature for the diffusion time between two nearest-neighbor octahedral sites of $\tau_{300 K}\simeq 10^{-9}$ s, i.e., a slower rate by one order of magnitude. A reduced value for the diffusion rate was likely, as for $x=1.1$ a double diffusion is necessary to move from one octahedral site to another (through the tetrahedral site), with a double barrier to pass. Yet, in the following, we shall rather consider the case where the deuterium ion is diffusing towards an octahedral site filled by a muonic deuterium, a neutral particule more akin to a point dipole than to a point charge (at the typical \AA~length-scale of electronic processes) and therefore, even for $x=1$, we expect the diffusion constant be rather of the order of the case where the octahedron is not filled, as the Coulomb repulsion of muonic deuterium can be neglected. Therefore we expect that the rate\cite{svare} $10^{10}$ s$^{-1}$ describes the process more correctly. We finally remind that diffusion data for deuterium\cite{svare,majorowski} show that deuterium mobility is higher than protium mobility (inverse isotope effect) in the temperature range 150K to 650 K, by a factor 1.5.
 
Once the deuterium ion tunnels to the same octahedral site of the deuterium muonic atom, the formation rate of the muonic $H_2^+$ molecule is higher than in the case of the H$_2$ atom, because there are no electrons to strip\cite{calcJack}: the deuterium ion approaches the muonic deuterium atom at thermal speeds ($\simeq 10^3$ m/s) and the latter has the time to polarize with the muon in between the two positive deuteria in a time negligible with respect to the $2\cdot 10^{-8}$ s estimated by Jackson. By readapting Jackson's calculations\cite{calcJack}, we might estimate this the molecule formation time smaller by a factor of about 500. 
If the above considerations are correct, it means that the formation time of the muonic H$_2^+$ molecule is less than the diffusion time ($t_d\simeq 10^{-10}$ s). So, $t_d$ controls the characteristic time of the molecular formation and $f_C$ in Eq. (\ref{mucycle}) can be as high as $10^{10}$ s$^{-1}$.

\subsection{Effective sticking probability: estimation of $f_{st}$ in Equaton (\ref{mucycle}).}

Once the muonic $H_2^+$ molecule is formed, the nuclear fusion process takes place at the same rate as calculated for 1957 {\it cold fusion}, as it is mainly a nuclear process mediated by the $\mu^-$ particle, therefore rather insensitive to the surrounding condensed matter properties\cite{note2}. 
The main action of the PdD crystal on the decrease of the sticking probability is associated to the increase of the reactivation process. In principle the muonic $^3\! {\rm He}^+$ can have a non-channeling or a channeling outgoing direction, but the latter case being rarer statistically, we neglect it in what follows.

In the case of a random direction that does not channel, we can evaluate the ionization rate through the same approach already used for non-cristalline deuterium\cite{rafelski}, only based on the average energy loss. In order to relate the initial sticking, $p_{is}$, to the final sticking $p_{st}$ after muon stripping, we consider the competition of the two processes: rate of energy loss of the muonic $^3\! {\rm He}^+$ in the PdD medium and rate of muon stripping, as the latter cannot take place once the muonic atom has stopped.\cite{notastop}
We have therefore the two equations:

\begin{subequations} 
\begin{align}
\hspace{0.5cm}\frac{dE}{dt}&=-\rho v S(E) \label{rafel11} \\
\hspace{0.5cm}\frac{dp_{st}}{dt}&\simeq -\sigma_{\rm str}(E)\rho v p_{st} \label{rafel12}
\end{align}
\label{rafel}
\end{subequations} 

\noindent where the stripping cross-section $\sigma_{\rm str}(E)$ is the sum of the ionization and transfer cross sections and $S(E)$ is the mass stopping power. By eliminating the time in the two equations, we obtain $\frac{dp_{st}}{p_{st}}=\frac{\sigma_{\rm str}(E)}{S(E)}dE$, with the final result:

\begin{align}
\hspace{1cm} p_{st} = p_{is} e^{-\int_{E_f}^{E_0}\frac{\sigma_{\rm str}(E)}{S(E)}dE}
\label{rafel2}
\end{align}

\noindent where $E_0$ is the initial energy (of the order of MeV) and $E_f\simeq 0$ the final energy.
In the hypothesis that $\sigma_{\rm str}(E)$ does not change much from liquid deuterium to PdD (and, in the case of changes, we argue that they would rather increase in PdD, because more stripping channels would set in, so that $p_{st}$ would be decreased), we show below that $p_{st}$ decreases in PdD, compared to liquid deuterium, because of the reduction of $S(E)$.

\begin{figure}[ht]
\centering
\includegraphics[width=0.4\textwidth]{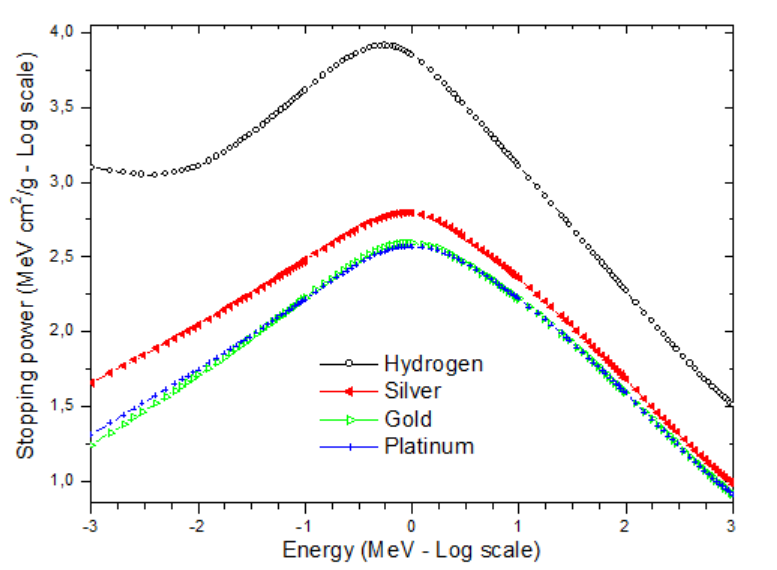}
\caption{Calculation with Astar of the mass stopping power $S(E)$ for $\alpha$ particles in the four materials: hydrogen, silver, platinum and gold (see text for explanations).}
\label{astar} 
\end{figure}

Our simulations are based on the Astar software\cite{nistastar,berger}, that for $\alpha$ particles uses the Ziegler, Biersack and Littmark approximation for the stopping power at low energy.\cite{zbl}  
Though $S(E)$ for PdD (or for Pd) is not available in the energy region from 1 keV to 1 MeV, $S(E)$ of Ag should be quite close to that of Pd: in general, the change from $nd^8$ to $nd^9$ elements on the same row should be limited. We checked for Pt and Au, where $S(E)$ was available\cite{nistastar}, see Fig. \ref{astar}, that they were practically the same. So, we used $S(E)$ of silver to mimic the behaviour of $S(E)$ in PdD and from Fig. \ref{astar} it is clear that, in the energy range of interest, $S(E)$ for silver is always more than one order of magnitude less than the $S(E)$ for hydrogen.

Whereas the linear stopping power $S^{\rm{(l)}}(E)$ of an alloy of two materials is given by their sum, in the case of the mass stopping power we get a weighted average: $S_{\rm PdD}(E)=\frac{1}{\rho_{\rm PdD}} S_{\rm PdD}^{\rm{(l)}}(E) = \frac{1}{\rho_{\rm PdD}} (S_{\rm Pd}^{\rm{(l)}}(E) + S_{\rm D}^{\rm{(l)}}(E))= \frac{\rho_{\rm Pd}}{\rho_{\rm PdD}}S_{\rm Pd}(E) + \frac{\rho_{\rm D}}{\rho_{\rm PdD}}S_{\rm D}(E)$, where $\rho_{\rm Pd}\simeq \rho_{\rm PdD} \simeq 11.3$ g/cm$^3$ and $\rho_{\rm D}\simeq \rho_{{\rm H}_2}^{(l)}\simeq 0.17$ g/cm$^3$ are the respective densities.

If we assume that $S_{\rm Pd}(E)\simeq S_{\rm Ag}(E)\leq 0.1 S_{\rm H}(E)$, for any energy below 1 MeV, then we get $S_{\rm PdD}(E) \simeq 0.115 S_{\rm H}(E)$, which, given equation (\ref{rafel2}), leads to $\frac{p_{st}^{\rm PdD}}{p_{st}^{\rm H}} \simeq \frac{e^{-9I}}{e^{-I}}$, where $I=\int_{E_f}^{E_0}\frac{\sigma_{\rm str}(E)}{S_{\rm H}(E)}dE$.

This means a reduction of $p_{st}$ for PdD compared to $p_{st}$ for hydrogen, of a value that depends on the unknown $I$ integral.
If we take a constant value (with energy) for $\sigma_{\rm str}(E)$ in the liquid hydrogen case, we can estimate it from the theoretical calculations\cite{ponomarev1} and get the rough value $\sigma_{\rm str}\simeq 10^2$ cm$^2$g$^{-1}$. We can then numerically average $S_{\rm H}(E)$ in the energy range [1 KeV, 1 MeV] from Fig \ref{astar}  and get $S_{\rm H}^{\rm (av)}= 5\cdot 10^2$ MeV cm$^2$/g. Finally, using $E_0-E_f=\Delta E \simeq 1$ MeV, we obtain $I\simeq \Delta E \cdot \sigma_{\rm str}/S \simeq 0.2$ and therefore $\frac{p_{st}^{\rm PdD}}{p_{st}^{\rm H}} \simeq \frac{e^{-1.8}}{e^{-0.2}}\simeq 0.2$.

So, $p_{st}$ would reduce by about $80\%$ in PdD compared to liquid hydrogen, i.e., for $dd$-fusion, from $p_{st}\simeq 0.12$ to $p_{st}\simeq 0.025$. However, this value should be considered {\it cum grano salis}, as the calculations performed here suffer from many not-fully-controlled approximations and only an experimental result can provide a reliable value.

\subsection{Cosmic $\mu^-$ implantation site in PdD crystal: guess of $f_X$ in Eq. (\ref{mucycle})}

The rate of the muon component in cosmic rays\cite{cosmicmu} (roughly 45\% of $\mu^-$ and 55\% of $\mu^+$) is around one per minute on a surface of the order of the cm$^2$. The most of the studies concern the branch of the energy-momentum spectrum with $p\ge 0.3$ MeV/c, as detailed in Ref. [\onlinecite{cosmicmu}]. In particular the most of muons are created around 15 km with an energy of 6 GeV and arrive at the sea level with 4 GeV (they loose around 2 MeV per g/cm$^2$ and the atmosphere interaction depth per vertical motion is around 1000 cm$^2$/g). Their energy distribution mainly depends on the angle of incidence and muons with a razing incidence angle can have an energy as low as few keV (using the stopping power of the atmosphere from Fig. 1 of Ref. [\onlinecite{cosmicmu}] and using a grazing path of $\simeq$300 km in the atmosphere). The density of PdD is 11.3 g/cm$^3$ and therefore an incident muon around its ionization minimum ($\simeq 1$ GeV) looses about 22 MeV in 1 cm, whereas a low-energy muon around 10 keV would stop in 10 nm (stopping power around 100 MeV g/cm$^2$ at this energy), i.e. less than three unit cells.

When a $\mu^-$ stops in PdD it can either be captured by a deuterium or by a palladium with a probablity that depends on the charge distribution within the unit cell and the energy of the muon. The most comprehensive review about muon atomic capture in solid crystals, Ref. [\onlinecite{physrep}] provides a list of experimental data and phenomenological laws about muon capture within several solid-state materials, though in all reported cases the high-Z atom wass the less electronegative - therefore positively charged - contrary to the case of PdD.
Self-consistent fdmnes calculations of the charge distribution in octahedral PdD gives a non-negligible weight ($\simeq 0.5$ electrons per PdD unit) in interstitial regions (e.g. tetrahedral sites), but even with a detailed knowledge of the electron-distribution in PdD, it is not easy to foresee where the $\mu^-$ particles would stop, for the following reasons. 

The dynamics of the negative muon depends on the energies at which it decays: at positive energies (where the energy zero is the limit of binding energy, i.e., about the work-function above the Fermi level of the crystal), the muon is still free to explore the whole crystal 'balistically': the energy levels form a continuum and the $\mu^-$ can decay to any energy, provided the transition-matrix elements are not zero. However, at 'small' negative (binding) energies, of the order of the Fermi energy, the muon would rather follow the states of the band-structure, with possible gaps for given $\vec{k}$ directions. It therefore acquires an effective band-structure mass given by $m^*=\left(\frac{1}{\hbar^2}\frac{\partial^2 \varepsilon_k}{\partial k^2} \right)^{-1}$, even if it is not constrained by the Pauli principle to fill empty electronic levels. Regardless, if it spends some femtoseconds in the PdD band-structure, before decaying to lower-lying energy levels, we cannot theoretically exclude the presence of a resonant transition to the bound deuterium $1s$-state. {\it Mutatis mutandis}, it would be the same idea used by Vesman\cite{vesman} and Ponomarev\cite{ponomarev1} to explain the resonant (fast) muonic molecular formation that could not be foreseen theoretically and had to be found experimentally.

It should be stressed that, in the framework of Eq. (\ref{mucycle}), it is not important when the first cosmic $\mu^-$ is captured by the $d^+$-ion. 
Of course, the probability of such a capture is related to the number of possible initial triggering events determined by cosmic muons. However, such a first capture only sets time $t=0$ and does not determine the rate of Eq. (\ref{mucycle}).
From this time $t=0$, in fact, we should evaluate the probability $p_d$ that the $\mu^-$ escaping from one of the two reactions (\ref{ddfus}), caracterized by keV-order-of-magnitude energy, is captured by the $d^+$-ion and not by the Pd nucleus. If we suppose that the $\mu^-$ can only be captured by the Pd-ion or by the $d^+$-ion, then we can call $p_X=1-p_d$ the probability that the $\mu^-$ is captured by the Pd nucleus. Given equation (\ref{mucycle}), we need that $p_X$ is not bigger than $p_{st}$ if we want to have a significant number $N_R$ of fusion reactions.

The precise estimate of the capture probability $p_X$ is fundamental for the final rate outcome and the deductions presented in this paper are not sufficiently solid as they should have been. For this reason, an experiment is suggested (see Section V), to be conducted at a muon source, to confirm (or infirm) the proposed picture.

\section{On the role of 'hot' neutrons to propagate deuterium fusion.}

In this section we discuss the recent theoretical\cite{prc1} and experimental\cite{prc2} demonstration of the role of 'hot' neutrons in propagating deuterium fusion, by reviewing the experimental results\cite{prc2} about neutron-induced deuterium-deuterium fusion in ErD$_3$ and TiD$_2$ and the possible role of the Pd lattice in PdD.

In a couple of recent papers\cite{prc1,prc2} Pines {\it et al.}\cite{prc1} have theoretically suggested and Steinetz {\it et al.}\cite{prc2} experimentally proven that deuterium-deuterium nuclear fusion can develop in a crystal matrix at ambient temperature, when a triggering event is introduced in the crystal. In the specific conditions analyzed in these papers, the triggering event was provided by $\gamma$-rays of energy slightly above the deuteron photodissociation energy ($E_{\gamma n}\simeq 2.23$ MeV). Photoneutrons with kinetic energy around 300 keV are produced and can scatter the deuterium fuel charged in the crystal matrix (ErD$_2$ or TiD$_3$ in their case). We remind that neutron scattering on deuterium is mainly elastic (with very modest absorption of $\simeq 0.5$ mb) and also that scattering of neutrons with kinetic energy lower than $\simeq 5$ MeV on heavy ions (palladium, but also erbium and titanium) is basically elastic (inelastic channels only open for higher neutron energies like those involved in the reaction of Eq. (\ref{dtfus})). Moreover, given their pronounced mass difference, neutrons impinging on palladium (erbium, titanium) ions loose less than $4\%$ of their kinetic energy, whereas the neutron/deuterium collision determines the transfer of up to $\frac{8}{9}$ of the neutron kinetic energy to the deuterium particle (in the case of zero impact parameter), the neutron remaining in this case with $\frac{1}{9}$ of its kinetic energy.
It was found in Ref. [\onlinecite{prc2}], in a six-hour measurement during which the photon irradiation was kept constant, that deuterium-deuterium collision produced about 3000 fusion events per second (1500 neutrons of 2.45 MeV per second were measured, covering only the neutron-helium channel of the $dd$ fusion, with 50\% branching ratio). Unfortunately, the rate of production of photoneutrons could not be measured, so that there is no precise estimation of the efficiency of the 300 keV neutron-deuteron scattering in producing fusion events. 

In the case of a Pd crystal, neutrons with maximal kinetic energy of 2.45 MeV are produced by the fusion process (\ref{ddfus2}) catalyzed by $\mu^-$. 
Each neutron can undertake at least two collisions with deuterium ions, with a kinetic energy above or around $\simeq 300$ keV: the first with the initial kinetic energy of 2.45 MeV, the second with the energy of at least $\frac{2.45}{9}\simeq 270$ keV left after the first collision. A deuterium with $\simeq 2$ MeV kinetic energy has a cross-section of about 0.1 barns for $dd$ fusion, the cross section reducing to $\simeq 0.05$ barns around 270 keV. If we consider that a neutron of 2.45 MeV kinetic energy has a speed of $\simeq 10^7$ m/s, then each neutron coming out of the reactions induced by the $\mu^-$ can impact a deuterium ion in a time from $2\cdot 10^{-17}$ s (if the impact is with the deuterium ion in the nearest-neighbor octahedral site, $\simeq 2$ \AA) to $10^{-15}$ s, i.e., practically instantaneously for the time-scales involved in the fusion reactions. A similar conclusion is true for the 270 keV neutron.
In the case of frontal collision between neutron and deuterium (null impact parameter), the deuterium leaves with $\frac{8}{9}$ of the kinetic energy, i.e., 2.18 MeV and 240 keV, respectively. The corresponding speeds are $\simeq 1.4\cdot 10^7$ m/s and $\simeq 5\cdot 10^6$ m/s, respectively. The corresponding fusion rates are therefore, using $f_r = \rho\sigma v$, where $\rho\simeq 11.3$ g/cm$^3$ is the density of the medium, $v$ the relative velocity, and $\sigma$ the cross section for the process under consideration, of $\simeq 10^7$ Hz and $\simeq 10^6$ Hz, respectively.

We do not think that it is realistic to perform a numerical simulation about the overall process, as we have no information about the damage and creation of defects in the crystal that a fusion process may determine locally in the PdD lattice. We can however try an estimate of the number of events needed to reach the Watt production measured in [\onlinecite{scara}]. If we consider the $10^7$ Hz rate estimated above, in order to get Watts in $dd$-fusion (\ref{ddfus2}), developing $4\cdot 10^6\cdot 1.6\cdot 10^{-19}\simeq 0.6 $ pJ energy per fusion reaction in a time $\Delta t\simeq 10^{-7}$ s, we obtain a power $P=6 \,\, \mu$W. We need therefore $10^6 \simeq 2^{20}$ processes of the kind (\ref{ddfus2}) that take place in parallel, if we want to reach 6 W, so a factor 20 of neutron doubling in the above chain reaction. Whether this factor is realistic or not, I think that only a dedicated experiment would tell us.

We conclude this Section reminding also the role possibly played by Oppenheimer-Phillips mechanism that might further reduce the Coulomb barrier by the polarization of the two colliding deuterons having the neutron part of the deuteron towards each other, a process that might be further stabilized by the electron distribution\cite{prc1,prc2}. We remark that this process reduces the spherical symmetry by selecting a special axis even in the center-of-mass frame (the axis of alignment of the neutron-proton direction) which might lead to a preferred direction of ejection of the fusion products (compared to the usual spherical symmetry of $dd$-fusion). This might explain a special feature of neutron-deuteron scattering in a crystal matrix, i.e., the distribution of angular scattering outcomes, that, compared to high pressure gases, seems to privilege smaller angles\cite{2013deutPd}. The latter point can also lead to many more scattering events, thereby implying a higher scattering rate compared to the liquid deuterium case, that might lead to the 20 doublings looked for in the chain reaction.

In summary, the main teaching of these two papers\cite{prc1,prc2} is that, under specific conditions (a constant external photon flux introducing some energy in the system), neutrons from $\simeq 100$ keV to a few MeV, produced by external agents, can trigger deuterium-deuterium nuclear-fusion reactions at ambient temperature. It remains to be seen if the overall mechanism suggested here can really be at the basis of 1989 cold fusion.

\section{Conclusions: A possible mechanism for 1989 cold fusion}

The present paper tries to explain some of the less contested facts associated to the 1989 {\it cold fusion} by coupling two, apparently far, experimental evidences: on one side, the $\mu^-$-catalyzed deuterium fusion\cite{alvarez,jackson1} at ambient temperature, well known since 1957, and on the other side the very recent experimental verification\cite{prc1,prc2} of the role played by 'hot' neutrons (kinetic energy around 300 keV) in a crystal matrix densely loaded with deuterium in triggering  deuterium fusion at ambient temperature. The key idea of the latter is that neutrons, as a function of their kinetic energy, might be extremely effective in allowing deuterium fusion by collision, even at ambient temperature, also playing on the electron screening of the deuterium-deuterium Coulomb repulsion (which can lead to the decrease of the $Ze^2/r$ Coulomb potential, by the constant factor $U_s$, of the order of 800 eV for palladium.\cite{terrasi}).

So, the picture that we propose is the following: cosmic $\mu^-$ with typical momentum less than few MeV/c stop in the PdD crystal and trigger several deuterium fusions (more than ten per muon) thereby freeing $\simeq 2.45$ MeV neutrons that, by collision can continue and propagate the fusion reaction. We remark that the role of the cosmic muon would just be to trigger the reaction, what could have in principle be done also by cosmic neutrons (by collision). The reason to propose the $\mu^-$ is that it is in principle much more efficient than a neutron to trigger fusion events (a cosmic neutron could trigger just one or maybe two fusion reactions by collision with deuterium ions, vs more than ten for the muon, in the most favorable conditions - see Table I below). As seen in Section III and IV, the peculiar features of the crystal lattice allow the propagation of the fusion reaction in a way that would not be possible for liquid fuel (like in a plasma), mainly because of the different collision conditions (e.g., the action of Pd nuclei as an elastic barrier for Mev and sub-MeV neutrons, changes the scattering conditions compared to liquid deuterium). In this way, for a short time, a chain reaction might set forward in the most favorable conditions. Whether this allows reaching the Watts of excess power for a few minutes that has been claimed\cite{scara}, it remains to be seen. This positive outcome would be anyhow strongly dependent on {\it alea}, thereby possibly explaining why several experiments in the past\cite{scara,cfno1,cfno2,cfno3} could not measure any excess power. The present mechanism also explains why the amount of 2.45 MeV neutrons is not necessarily correlated with the amount of excess heat and usually much less than what would be expected on the basis of Eq. (\ref{ddfus2}) - both because of the three-body reaction of Eq. (\ref{ddfus2}) and because a large part of these neutrons is involved in the scattering with $d^+$-ions, thereby loosing its kinetic energy. The big amount of internal collisions in PdD, compared to D$_2$ high-pressure gas, might also explain why the excess power is rather correlated\cite{cfyes2} with $^4$He than $^3$He, part of the latter being eventually transformed to $^4$He through the reaction $d^+ + ^3\! He^{++} \rightarrow ^4\! He^{++} + p^+  (+ Q_4)$, with $Q_4 \simeq 18.3$ MeV or by reaction (\ref{dtfus}). Finally, we should not forget that, though we mainly focused on reaction (\ref{ddfus2}), also reaction (\ref{ddfus1}) can take place with $50\%$ probability. In this case, the two outgoing $p^+$ and $t^+$ particles are singly charged, so that the sticking probability is less effective (though not measured in the literature\cite{petitjean}). Actually, if the $\mu^-$ was sticked to the $t^+$, the reaction would continue with a higher rate and energy emitted for the $t\mu$ atom, through reaction (\ref{dtfus}), and with a much lower rate\cite{jackson1,jackson2} and energy emitted for the $p\mu$ atom. In any case, also for Eq. (\ref{ddfus1}), the critical parameter would be $p_X$.

\begin{table}[ht]
\begin{center}
\caption{Parameters of Eq. (\ref{mucycle}): estimation for PdD vs. liquid deuterium experimental values}
\begin{tabular}{ |c|c|c|c|c| } 
 \hline
  & $f_c$ & $p_{st}$ & $p_X$ & $N_R$ \\ 
 \hline
 liquid deuterium & $10^8$ Hz & $12\%$ & 0 & 8 \\ 
 \hline
 PdD ($^{\mu}$He$_3\simeq 1 MeV$)  & $10^{10}$ & $2.4\%$ & $1\%$ to $10\%$ & 30 to 8 \\ 
 \hline
\end{tabular}
\end{center}
\end{table}

So, our conclusions are strongly dependent on two educated guesses advanced in this paper on the basis of the present theoretical knowledge, that should be checked experimentally: 1) the efficiency of one $\mu^-$ to drive a few tens of fusion reactions with purely deuterium fuel embedded in the palladium crystal, which critically depends on the unknown parameter $p_X$, and 2) the efficiency of 'hot' (300 keV to 2 MeV) neutrons to continue the fusion after the initial $\mu^-$-triggering, to eventually reach the Watts power. Both 1) and 2) might be experimentally tested at a muon source with the following experiment on a PdD crystal. First, prepare an electrolithic cell with a calorimeter following the prescriptions for the choice of the Pd cathod,\cite{scara,cfyes1,cfyes2} shielded from cosmic rays. Then send a variable-intensity $\mu^-$ beam on the crystal, looking for fusions outcomes (as it was done to test 1957 {\it cold fusion}, for example at PSI\cite{petitjean}). Check for the optimal energy at which the muons should be sent on the sample in order to produce fusion events. Reduce the muon flux to measure the minimum flux inducing fusion reactions and then stop the $\mu^-$ flux to monitor whether fusion events continue even in its absence, catalyzed by the 'hot' neutrons produced by the previous, $\mu^-$-catalyzed, fusions. An eventual measurement of excess heat in this case, besides testing the effectiveness of the present ideas could also point to the best conditions, in terms of initial muon energy, to have an optimal rate, and might open a new field of researches. Of course, the absence of any signal under intense $\mu^-$-flux would point to a too high value of $p_X$ and therefore definitely invalidate the proposed mechanism.

{\bf{Acknowledgments}} This paper would have never been written had I not met late F. Scaramuzzi, who convinced me of the reality of some heat production in his deuterium-loaded palladium cells. I also thank C.R. Natoli for several discussions on the subject and A. Telese for useful hints about triggers and burnings.


\vspace{0.5cm}


\begin{thebibliography}{99}
\bibitem{alvarez}
L.W Alvarez, H. Bradner, F.S. Crawford Jr., J.A. Crawford, P. Falk-Vairant, M.L. Good, J.D. Gow, A.H. Rosenfeld, F. Solmitz, M.L. Stevenson, H.K. Ticho, and R.D. Tripp, 'Catalysis of nuclear reactions by $\mu$ mesons' Phys. Rev. {\bf 105}, 1127-1128 (1957)
\bibitem{jackson1}
J.D. Jackson, 'Catalysis of nuclear reactions between hydrogen isotopes by $\mu^-$ mesons', Phys. Rev. {\bf 106}, 330-339 (1957) 
\bibitem{jackson2}
J.D. Jackson, 'A personal adventure in muon-catalyzed fusion', Phys. Perspect. {\bf 12}, 74-88 (2010)
\bibitem{prlexp1987}
H. Bossy, H. Daniel, F. J. Hartmann, W. Neumann, H.S. Plendl, G. Schmidt, T. von Egidy, W.H. Breunlich, M. Cargnelli, P. Kammel, J. Marton, N. Nagele, A. Scrinzi, J. Werner, J. Zmeskal, and C. Petitjean, 'Determination of Muonic Helium X-Ray Yields after Muon-Catalyzed pd, dd, and dt Fusion', Physical Review Letters {\bf 59}, 2864-2867 (1987)
\bibitem{aipconfproc}
A whole volume of the AIP Conf. Proc. was dedicated to the subject: Vol. {\bf 181}, (1988). See, in particular, J. Rafelski, 'The challenges of muon catalyzed fusion' AIP Conf. Proc. {\bf 181}, 451–464 (1988)
\bibitem{petitjean}
D.V. Balin, V.A. Ganzha, S.M. Kozlov, E.M. Maev, G.E. Petrov, M.A. Soroka, G.N. Schapkin, G.G. Semenchuk, V.A. Trofimov, A.A. Vasiliev, A.A. Vorobyov, N.I. Voropaev, C. Petitjean, B. Gartner, B. Lauss, J. Marton, J. Zmeskal, T. Case, K.M. Crowe, P. Kammel, F.J. Hartmann, M.P. Faifman, 'High Precision Study of Muon Catalyzed Fusion in D$_2$ and HD Gas' Physics of Particles and Nuclei {\bf 42}, pages 185–214, (2011) [DOI: 10.1134/S106377961102002X]
\bibitem{pons1}
M. Fleischmann, S. Pons and M. Hawkins, 'Electrochemically induced nuclear fusion of deuterium', J. Electroanal. Chem. {\bf 261} (1989) 301–308. Erratum in {\bf 263} (1989) 187–188.
\bibitem{pons2}
M. Fleischmann, S. Pons, M.W. Anderson, L.J. Li and M. Hawkins, 'Calorimetry of the palladium-deuterium-heavy water system', J. Electroanal. Chem. {\bf 287} (1990) 293–348
\bibitem{petrasso}
R.D. Petrasso, X. Chen, K.W. Wenzel, R.R. Parker, C.K. Li, C. Fiore, 'Measurement of $\gamma$-rays from cold fusion', Nature {\bf 339}, 183-185 (1989); and erratum, {\bf 339}, 264 (1989).
{\it Comment} M. Fleischmann, S. Pons, M. Hawkins and R.J. Hoffman, Nature {\bf 339}, 667 (1989). 
{\it Reply to Comment} R.D. Petrasso, X. Chen, K.W. Wenzel, R.R. Parker, C.K. Li, C. Fiore, Nature {\bf 339}, 667-669 (1989)
\bibitem{salamon}
M.H. Salamon, M.E. Wrennt, H.E. Bergeson, K.C. Crawford, W.H. Delaney, C.L. Hendersontt, Y.Q. Li, J.A. Rusho, G.M. Sandquist, and S.M. Seltzer, 'Limits on the emission of neutrons, $\gamma$-rays, electrons and protons from
Pons/Fieischmann electrolytic cells', Nature {\bf 344}, 401-405 (1990)
\bibitem{vesman}
E.A. Vesman, Pis'ma Zh. Eksp Theor. Fiz. {\bf 5}, 113 (1967) [JETP Lett. {\bf 5}, 91 (1967)]. 
\bibitem{ponomarev1}
L.I. Ponomarev, I.V. Puzyinin and T.P. Puzyinina, Comput. Phys. {\bf 13}, 1 (1973). S.I. Vinitsky, L.I. Ponomarev, I.V. Puzyinin {\it et al.}, Zh. Eksp. Theor. Fiz. {\bf 74}, 849 (1978)
\bibitem{ponomarev2}
L.I. Men'shikov and L.I. Ponomarev, 'Effect of the reaction $d\mu (n)+t\rightarrow d+t\mu (n)$ on the kinetics of muon-catalysis processes in a  $D_2 + T_2$ mixture.' Pis'ma Zh. Eksp Theor. Fiz. {\bf 39}, 542-545 (1984) [JETP Lett. {\bf 39}, 663-667 (1984)]. 
L.I. Men'shikov and L.I. Ponomarev, 'Charge exchange of excited mesic atoms of hydrogen isotopes in triple collisions with molecules' Pis'ma Zh. Eksp Theor. Fiz. {\bf 42}, 12-14 (1985) [JETP Lett. {\bf 42}, 13-16 (1985)]
\bibitem{note1}
Moreover, the errors in the energy-scale identifications of published papers, like in the Comment to [\onlinecite{petrasso}], provided a feeling of methodological incompetence.
\bibitem{scara}
COLD FUSION: The history of research in Italy, Editors: Sergio Martellucci, Angela Rosati, Francesco Scaramuzzi, Vittorio Violante (2008), ISBN 978-88-8286-204-6, https://www.lenr-canr.org/acrobat/ENEAcoldfusion.pdf
\bibitem{rmp-jap}
Setsuo Ichimaru, 'Nuclear fusion in dense plasmas' Rev. Mod. Phys. {\bf 65}, 255-299 (1993)
\bibitem{cfyes1}
M.C.H. McKubre, S. Crouch-Baker, R.C. Rocha-Filho, S.I. Smedley, F.L. Tanzella, T.O. Passell and J. Santucci, 'Isothermal Flow Calorimetric Investigations of the D/Pd and H/Pd Systems', J. Electroanal. Chem. {\bf 368}, (1994) 55-71.
\bibitem{cfyes2}
B.F. Bush, J.J. Lagowski, M.H. Miles and G.S. Ostrom, 'Helium production during the electrolysis of D2O in cold fusion experiments' J. Electroanal. Chem. {\bf 304} (1991) 271-278
\bibitem{cfno1}
J.F. Ziegler, T.H. Zabel, J.J. Cuomo, V.A. Brusic, G.S. Cargill, III, E.J. O'Sullivan and A.D. Marwick, 'Electrochemical Experiments in Cold Nuclear Fusion', Phys. Rev. Lett. {\bf 62}, (1989), 2929-2932
\bibitem{cfno2}
Peter M. Richards, 'Molecular-dynamics investigation of deuteron separation in PdD$_{1.1}$', Phys. Rev. B {\bf 40}, (1989) 7966-7968
\bibitem{cfno3}
S.M. Myers, P.M. Richards, D.M. Follstaedt, and J.E. Schirber, 'Superstoichiometry, accelerated diffusion, and nuclear reactions 1n deuterium-implanted palladium', , Physical Review B {\bf 43}, 9503-9510 (1991)
\bibitem{terrasi}
F. Raiola, P. Migliardi, L. Gang, C. Bonomo, G. Gyürky, R. Bonetti, C. Broggini, N.E. Christensen, P. Corvisiero, J. Cruz, A. D’Onofrio, Z. Fülöp, G. Gervino, L. Gialanella, A.P. Jesus, M. Junker, K. Langanke, P. Prati, V. Roca, C. Rolfs, M. Romano, E. Somorjai, F. Strieder, A. Svane, F. Terrasi, J. Winter, 'Electron screening in $d(d, p)t$ for deuterated metals and the periodic table', Physics Letters B {\bf 547}, 193–199 (2002)
\bibitem{diffdD}
In this paper we use for deuterium both the chemist convention to use capital $D$ (namely, in molecules) and the nuclear physicist convention to use the lower-case $d$ (namely, for nuclear reactions)
\bibitem{notamu}
However, this flux is only measured for muons with quantity of movement $p\ge 3$ MeV/c. No clear data exist for lower-energy muons, that are probably most effective in binding to deuterium ions.
\bibitem{prc1}
V. Pines, M. Pines, A. Chait, B.M. Steinetz, L.P. Forsley, R.C. Hendricks, G.C. Fralick, T.L. Benyo, B. Baramsai, P.B. Ugorowski, M.D. Becks, R.E. Martin, N. Penney, and C.E. Sandifer, 'Nuclear fusion reactions in deuterated metals', Phys. Rev. C {\bf 101}, 044609-1 to 12 (2020)
\bibitem{prc2}
B.M. Steinetz, T.L. Benyo, A. Chait, R.C. Hendricks, L.P. Forsley, B. Baramsai, P.B. Ugorowski, M.D. Becks, V. Pines, M. Pines, R.E. Martin, N. Penney, G.C. Fralick, and C.E. Sandifer,'Novel nuclear reactions observed in bremsstrahlung-irradiated deuterated metals', Phys. Rev. C {\bf 101}, 044610-1 to 13 (2020)
\bibitem{jackson1b}
See note 4 of Ref. [\onlinecite{jackson1}].
\bibitem{japreactivation}
Yoshiharu Mori, 'Enforced stripping of negative muons from $\mu$He$^+$ ions to stimulate muon-catalyzed fusion by cyclotron resonance acceleration', Prog. Theor. Exp. Phys. {\bf 2021}, 093G01-18 pages (2021), DOI: 10.1093/ptep/ptab111
\bibitem{kittel}
Charles Kittel 'Introduction to Solid State Physics' (2$^{\rm nd}$ edition), John Wiley $\&$ Sons (1956), pages 332-335
\bibitem{hydrogen}
In this section we employ the label H or the word hydrogen, for any isotope ($p$, $d$, or $t$), as in condensed-matter processes, related to the electronic configuration, there is no relevant difference between them.
\bibitem{nmr}
R.E. Norberg, 'Nuclear magnetic resonance of hydrogen absorbed into palladium wires', Physical Review {\bf 86}, 745-752 (1952)
\bibitem{totalenergy}
A.C. Switendick, 'Electronic structure and stability of palladium hydrogen (deuterium) systems, PdH(D)$_n$, $1\leq n\leq 3$', J. Less-Common Met. {\bf 172-174}, 1363-1370 (1991)
\bibitem{papacon}
D.A. Papaconstantopoulos, B.M. Klein, E.N. Economou, L.L. Boyer, 'Band structure and superconductivity of PdD$_x$ and PdH$_x$' Physical Review B {\bf 17}, 141-150 (1978)
\bibitem{fdmnes}
Y. Joly, "X-ray absorption near-edge structure calculations beyond the muffin-tin approximation". Physical Review B. {\bf 63} 125120 (2001); FDMNES project: https://fdmnes.neel.cnrs.fr/
\bibitem{resistivity}
J.P. Burger, S. Senoussi and B. Soufachi, 'Electrical and magnetic properties of Palladium Hydrides compared with those of pure Palladium', J. Less-Common Met. {\bf 49}, 213-222 (1976)
\bibitem{neutrondiff}
J.E. Worsham Jr., M.K. Wilkinson, and C.G. Shull, 'Neutron-diffraction observations on the palladium-hydrogen and palladium-deuterium systems', J. Phys. Chem. Solids {\bf 3}, 303-310 (1957)
\bibitem{svare}
I. Svare, 'Hydrogen diffusion by tunneling in Palladium', Physica B {\bf 141}, 270-276 (1986)
\bibitem{russians}
S.A. Semiletov, R.V. Baranova, Yu.P. Khodyrev, and R.M. Imamov, 'Electron-diffraction investigation of tetragonal PdH$_{1.33}$', Sov. Phys. Crystallogr. {\bf 25}, 665-669 (1980) [Kristallografiya {\bf 25}, 1162-1168]
\bibitem{kuji}
T Kuji, W.A. Oates, B.S. Bowerman and T.B. Flanagan, 'The partial excess thermodynamic properties of hydrogen in palladium', J. Phys. F: Met. Phys. {\bf 13}, 1785-1800 (1983)
\bibitem{nasa}
D.A. Otterson and R.J. Smith, 'Absorption of Hydrogen by Palladium and electrical resistivity up to Hydrogen-Palladium atom ratios of 0.97', NASA technical note NASA TN D-5441, {\tiny https://ntrs.nasa.gov/api/citations/19690027894/downloads/19690027894.pdf}
\bibitem{prl3vuoti}
N. Lopez, Z. Lodziana, F. Illas and M. Salmeron, 'When Langmuir is too simple: $H_2$ dissociation on Pd(111) at high coverage', Phys. Rev. Lett. {\bf 93}, 146103-1 to 4 (2004)
\bibitem{note1b}
We remark that, in stochiometric $PdD$, the $D$ density is 6$\cdot$10$^{22}$ d/cm$^3$, about a factor 1.5 bigger than in liquid hydrogen. Though it can be made bigger by filling also tetrahedral sites\cite{cfno3}, this would not necessarily increase the formation rate of muonic $D_2^+$, that critically depends also on the diffusion rate, the latter depending on the filling in a non monotonous way.
\bibitem{zunger}
Su-Huei Wei and A. Zunger, 'Stability of atomic and diatomic hydrogen in FCC palladium', Solid State Comm. {\bf 73} 327-330 (1990)
\bibitem{jpsj}
S. Ichimaru, S. Ogata, and A. Nakano, 'Rates of nuclear fusions in metal hydrides', Jour. Phys. Soc. Jpn. {\bf 59}, 3904-3915 (1990)
\bibitem{majorowski}
S. Majorowski and B. Baranowski, J. Phys. Chem. Solids {\bf 43}, 1119- (1982)
\bibitem{calcJack}
The details of Jackson's calculation on molecule formation (section D of Ref. [\onlinecite{jackson1}]) show that the stripping of the atomic electron, evaluated in terms of the transition matrix element of its wavefunction with the muon wave-function via the Coulomb repulsion, gives a factor of about 1/500 to the rate. In the absence of the electron, such a factor becomes unitary. Though the actual process in Pd crystal should not be described in the same framework, we may, in a first approximation suppose that the rate is roughly a factor 500 bigger than evaluated by Jackson.
\bibitem{note2}
Actually, an influence of condensed-matter properties on the nuclear process (\ref{ddfus2}) exists: the initial sticking probability depends on the state of nuclear spin of the muonic $^{\mu}\!D_2^+$ molecule, that partly changes the statistical ratio of Eqs. (\ref{ddfus1}) and (\ref{ddfus2}), up to a factor 1.5 (see Ref. (\onlinecite{petitjean}) and Refs. therein). In fact, if we consider the resonant $(J=1,\nu=1)$ state, with less than 2 eV of binding energy (here $J$ is the total angular momentum of the two deuterium ions, $\nu$ their vibrational state), its coupling to the muon can take place in the hyperfine $F=3/2$ or $F=1/2$ states, whose energy difference is $\simeq 48$ meV. Therefore, not only temperature changes around ambient temperature can influence the statistical ratio of Eqs. (\ref{ddfus1}) and (\ref{ddfus2}), but - we argue - also the local ferromagnetic fluctuations of Pd ion through direct exchange coupling can change it. 
In fact, it is known that pure Pd is 'almost' ferromagnetic,\cite{burger} in the sense that its high density of states at the Fermi level (mainly due to 4d-orbitals, as shown in Fig. \ref{fig1}) leads to 'almost' fulfill the Stoner criterion for ferromagnetism: it is sufficient to add less than 0.1\% of Fe or Co impurities to switch to a ferromagnetic state. Even though hydrogenation reduces its magnetic properties (as H electrons fill palladium 4d-orbitals, magnetic susceptibility goes to zero\cite{kittel,chizero} in PdH$_x$ for $x\ge 0.65$ and local magnetic moments become very weak, with diamagnetic coupling), ferromagnetic coupling of local magnetic moments can be restored\cite{burger} by doping the palladium crystal with Fe or Co up to 10\% before hydrogenating it.\cite{pdcomagmom} In the hypothesis that local magnetic moments can be kept up to $x=1$ (experiments\cite{burger} were only performed for PdH$_x$ up to $x=0.75$), such moments, through direct exchange coupling, would induce a ferromagnetic coupling on nuclear spins of neighbors  muonic $^{\mu}\!D_2^+$ molecule, thereby forcing the hyperfine $F=3/2$ state with a variation of the initial sticking probability. 
\bibitem{burger}
J.P. Burger, S. Senoussi and B. Soufach\'e, 'Electrical and magnetic properties of Palladium hydrides compared with those of pure Palladium', J. Less-Comm. Met., {\bf 49}, 213-222 (1976)
\bibitem{chizero}
R.J. Miller, T. Brun, and C.B. Satterthwaite, 'Magnetic susceptibility of Pd-H and Pd-D at temperatures between 6 and 150 K' Phys. Rev. B {\bf 18}, 5054-5059 (1978) 
\bibitem{pdcomagmom}
S. Akamarua, A. Kimurab, M. Haraa, K. Nishimurab, T. Abea, 'Hydrogenation effect on magnetic properties of Pd–Co alloys', J. Mag. Mag. Mat. {\bf 484}, 8-13 (2019)
\bibitem{rafelski}
J. Rafelski and D. Harley, 'Muon Catalyzed Fusion at High Density', Particle Accelerators {\bf 37-38},  409-416  (1992)
\bibitem{notastop}
The stopping of the muonic molecule with an initial energy around a MeV takes place in a much less time than either the muon lifetime $\tau_{\mu}$ or the inverse cycling rate $f_C^{-1}$.
\bibitem{nistastar}
The Astar software can be found at {\small https://physics.nist.gov/PhysRefData/Star/Text/ASTAR.html}
\bibitem{berger}
M.J. Berger, 'ESTAR, PSTAR, ASTAR: A PC package for calculating stopping powers and ranges of electrons, protons and helium ions - Version 2' U.S. National Institute of Standards and Technology: https://www-nds.iaea.org/publications/iaea-nds/iaea-nds-0144.pdf
\bibitem{zbl}
J.F. Ziegler, J.P. Biersack, and U. Littmark. In The Stopping and Range of Ions in Matter, volume 1, New York, 1985. Pergamon. ISBN 0-08-022053-3   
\bibitem{cosmicmu}
D.E. Groom, N.V. Mokhov, and S.I. Striganov, 'Muon stopping power and range tables 10 MeV to 100 TeV', LBNL-44742, Atomic Data and Nuclear Data Tables, Vol. {\bf 76}, No. 2, July 2001, p. 1-37 
\bibitem{physrep}
D.F. Measday, 'The nuclear physics of muon capture', Physics Report {\bf 354}, (2001) 243–409
\bibitem{kulsrud}
R.M. Kulsrud, 'A proposed method for reducing the sticking constant in muon cold fusion', AIP Conf. Proc. {\bf 181}, 367–380 (1988) https://doi.org/10.1063/1.37877
\bibitem{rafelskimu}
H.E. Rafelski, B. Moller, J. Rafelski, D. Trautmann and R.D. Viollier, 'Muon Reactivation in Muon-Catalyzed D-T Fusion', Prog. Part. and Nucl. Phys. {\bf 22}, (1989), Pag. 279-338
\bibitem{ddelscreening}
U. Greife, F. Gorris, M. Junker, C. Rolf, and D. Zahnow, 'Oppenheimer-Phillips effect and electron screeingn in d-d fusion reactions', Z. Phys. A {\bf 351}, 107-112 (1995)
\bibitem{neutdeut}
J.D. Seagrave and R.L. Henkel, 'Total cross-section of deuterium for neutrons from 0.2 to 22 MeV', Phys. Rev. {\bf 98}, 666-669 (1955)
\bibitem{pdelscreening}
A. Huke, K. Czerski, P. Heide, G. Ruprecht, N. Targosz, and W. Zebrowski, 'Enhancement of deuterium fusion reactions in metals and experimental implications', Phys. Rev. C {\bf 78}, 015803-1 to 20 (2008)
\bibitem{2013deutPd}
A.Yu. Didyka and R. Wi\'sniewski, 'Phenomenological Nuclear-Reaction Description in Deuterium-Saturated Palladium and Synthesized Structure in Dense Deuterium Gas under $\gamma$-Quanta Irradiation', Physics of Particles and Nuclei Letters {\bf 10} (2013), 273–287.
\bibitem{pdneutcrosssec}
K. Sameyoshi, and K. Abe, '14.8 MeV neutron induced activation in some metals and practical alloys', J. Nucl. Mat. {\bf 133-134}, 902-906 (1985)
\end{thebibliography}
\end{document}